\documentclass[aps,prx,floatfix,groupedaddress,nofootinbib,notitlepage,showpacs,superscriptaddress,twocolumn]{revtex4-1}
\usepackage{amsmath}
\usepackage{times,bm,bbm,bbold,graphicx,graphics,amssymb,amsmath,amsfonts,dsfont,hyperref,color}
\usepackage{mathrsfs,cancel}
\usepackage[utf8]{inputenc}
\usepackage{graphicx}
\usepackage{soul,xcolor}
\setstcolor{red}
\definecolor{mycolor}{rgb}{0.1, 0.1, 0.7}
\usepackage{appendix}
\usepackage{dsfont}
\usepackage{url}
\usepackage{soul}
\usepackage{siunitx}
\usepackage{physics}
\usepackage{mathtools}
\DeclareFontFamily{OT1}{pzc}{}
\DeclareFontShape{OT1}{pzc}{m}{it}
{<-> s * [1.25] pzcmi7t}{}
\DeclareMathAlphabet{\mathpzc}{OT1}{pzc}
{m}{it}
\hypersetup{
	plainpages=true,
	breaklinks=true,
	hypertexnames=false,
	pageanchor=true,
	colorlinks=true,
	linkcolor={mycolor},
	citecolor={mycolor},
	urlcolor={mycolor},
	anchorcolor={black}
}

\begin{document}
\title{The Multimode Character of Quantum States Released from a Superconducting Cavity}

\author{Maryam Khanahmadi}
\email{m.khanahmadi@chalmers.se}
\affiliation{Department of Microtechnology and Nanoscience, Chalmers University of Technology, 412 96 Gothenburg, Sweden}
\author{Mads Middelhede Lund}
\affiliation{Center for Complex Quantum Systems, Department of Physics and Astronomy, Aarhus University, 8000 Aarhus C, Denmark} 
\author{Klaus M{\o}lmer}
\affiliation{Niels Bohr Institute, University of Copenhagen, Blegdamsvej 17, DK-2100 Copenhagen,Denmark} 
\author{Göran Johansson}
\email{goran.l.johansson@chalmers.se}
\affiliation{Department of Microtechnology and Nanoscience, Chalmers University of Technology, 412 96 Gothenburg, Sweden}

\begin{abstract}

Quantum state transfer by propagating wave packets of electromagnetic radiation requires tunable couplings between the sending and receiving quantum systems and the propagation channel or waveguide. The highest fidelity of state transfer in experimental demonstrations so far has been in superconducting circuits. Here, the tunability always comes together with nonlinear interactions, arising from the same Josephson junctions that enable the tunability. The resulting non-linear dynamics correlates the photon number and spatio-temporal degrees of freedom and leads to a multi-mode output state, for any multi-photon state. In this work, we study as a generic example the release of complex quantum states from a superconducting resonator, employing a flux tunable coupler to engineer and control the release process.
%The resulting hamiltonian also describes the dynamics of recent experiments.  
We quantify the multi-mode character of the output state and discuss how to optimize the fidelity of a quantum state transfer process with this in mind.
\end{abstract}

\maketitle

\section{Introduction}

The exchange of quantum states  between distant locations is an important ingredient in secure communication networks and in scalable architectures for quantum computing \cite{kimble2008quantum,wehner2018quantum}. 
%Surface or bulk acoustic waves and 
%Electromagnetic waves form bosonic fields well suited for this purpose, as the corresponding quantized harmonic oscillator modes may contain multiple phonons or photons and thus faithfully map the state of multi-level emitters.
Quantum bits encoded in the higher dimensional oscillator modes of superconducting cavities have been demonstrated to withstand photon losses and permit elementary error correction \cite{vlastakis2013deterministically,leghtas2015confining,ofek2016extending,campagne2020quantum,puri2020bias,ni2023beating}. It would be desirable to use such multiphoton quantum states also for quantum communication purposes \cite{grimsmo2020quantum,PhysRevA.106.042614}.

While a linear mapping between a single oscillator mode and the continuum of propagating field modes, in principle, transfers the quantum state of the former to a traveling single-mode pulse, the temporal control of the release process is not trivial. In superconducting circuits, tunable couplers based on Josephson junctions are employed to control the evolution and release process in different architectures, such as fixed-frequency transmons, flux tunable transmons, or tunable transmission line resonators  \cite{houck2007generating,pierre2014storage, PhysRevX.4.041010,PhysRevA.93.063823,PhysRevApplied.8.054015,pfaff2017controlled,axline2018demand,cozzolino2019high,burkhart2021error,yang2023deterministic}. While the non-linearity of the Josephson junction enables tunable coupling, it also adds effective self-Kerr and cross-Kerr terms to the oscillator Hamiltonian. These non-linear terms may entangle the spatio-temporal release with the photon number contents of the pulse, and thus the emission becomes multi-mode in character and it may not function properly in a quantum network. 

In this article, we present a general analysis that takes the multi-mode character of the emission process fully into account. We employ a master equation approach that readily incorporates both the coherent coupling to the output field and decay and decoherence channels, and we use the quantum regression theorem to assess the mode decomposition of the emitted radiation. 

For the more quantitative discussion, we consider superconducting circuits. With low loss rates and strong coupling, these are promising platforms to efficiently prepare and emit quantum states into propagating modes \cite{krantz2019quantum,blais2021circuit}. Different studies and experiments have been done with a low number of photons
\cite{yin2013catch,pechal2014microwave}. In this paper, we theoretically analyze an experimentally relevant superconducting circuit architecture for which we can control the out-coupling strength and compute the accompanying non-linear couplings. The propagation transfer and the recapture of the field by downstream circuit components can then be analyzed by the method presented in  \cite{kiilerich2019input}.

The article is structured as follows:
In Sec. \ref{Theory}, we provide the formalism determining the characteristics of the output field of the quantum system.
In Sec. \ref{SLR}, we describe the superconducting emitter and tunable out-coupler in detail, and in Sec. \ref{sec:results}  we provide numerical results and study the quantitative effects of the nonlinearity on the multi-mode character of different  bosonic quantum states released from the circuit. Finally, we summarize the paper in Sec. \ref{Sum}. 

\section{Multimode Theory}\label{Theory}
\begin{figure*}[ht!]
\includegraphics[width=1\textwidth]{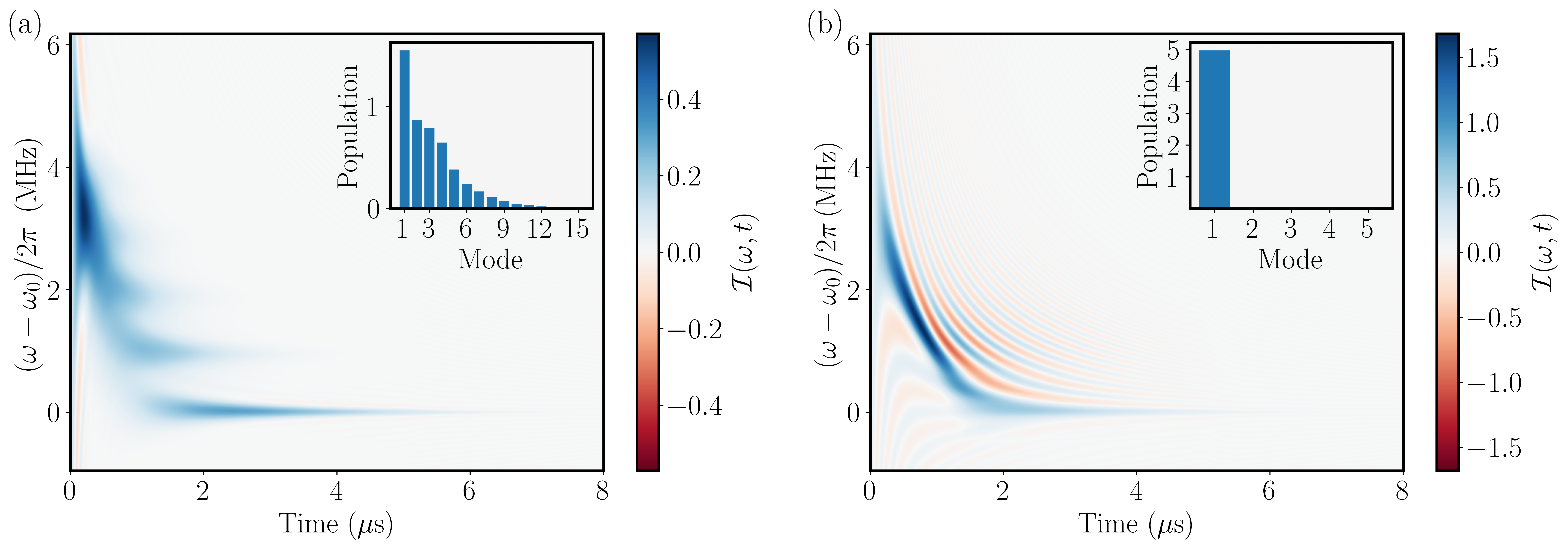}
\caption{Time-dependent spectra of the output field from a non-linear (a) and a linear (b) resonator. Panel (a) shows the  radiation from a non-linear resonator with constant frequency $\omega_0$, outcoupling rate $1/\kappa = 1 \,\mu s $, and Kerr coefficient $\chi/2\pi = .47 $ MHz prepared in an initial Fock state $\ket{n} = \ket{5}$. The insert in panel (a) shows that the emitted radiation occupies several eigenmodes.  In panels (b) we assume a vanishing Kerr coefficient $\chi=0$ and a time-dependent oscillator frequency $\omega(t)=\omega_0 + $max$(0,2\chi(n\exp(-\kappa t)-1))$. As shown in the inset of panel (b), the output field is a single-mode Fock state with $n=5$.} 
\label{Fig1}
\end{figure*}

In this section, we consider a single nonlinear resonator as a toy model for a non-linear emitter with the Hamiltonian 
\begin{align}\label{S1}
H_S(t) = \omega(t)a^\dagger a+ \chi(t) a^{\dagger 2}a^2,
\end{align}
having time-dependent frequency $\omega(t)$ and self-Kerr coefficient $\chi(t)$, using angular frequency units of energy ($\hbar =1$).
If we assume the emitter is coupled to a waveguide with constant strength $\sqrt{\kappa}$, the evolution of the reduced density matrix $\varrho(t)$ of the resonator is described by the Lindblad master equation with a single Lindblad operator $\sqrt{\kappa}$, describing the dissipation to the waveguide,
\begin{align}\label{me}
    \frac{\partial \varrho(t)}{\partial t} = -i[H_S(t), \varrho(t)] + \kappa\Bigg(a \varrho(t)a^\dagger - \frac{1}{2}[a^\dagger a, \varrho(t)]\Bigg).
\end{align} 
%In our simulation, we eliminate the effect of extra dissipation channels; $\kappa_i =0, i=1,..,n$, however, in the experiment it is important to be taken this into account. 
To find the mode decomposition of the emitted field one can utilize its first-order autocorrelation function \begin{align}\label{g12}
\mathcal{G}^{(1)}(t_1,t_2) =  \kappa\langle a^\dagger(t_2)a(t_1)\rangle = \sum_i n_i v^*_i(t_1)v_i(t_2),
\end{align}
which can be expressed in terms of the orthonormal output modes $v_i(t)$, with the mean photon number $n_i$.
The autocorrelation function $\mathcal{G}^{(1)}(t_1,t_2)$ can be calculated using the quantum regression theorem \cite{breuer2002theory,gardiner2004quantum}
\begin{align}\label{g1}
\mathcal{G}^{(1)}(t_1,t_2) = \kappa\mathrm{Tr}\Bigg[ a^\dagger \mathcal{L}(t_2,t_1) [a\mathcal{L}(t_1,0)\varrho_s(0)]\Bigg],
\end{align}
where $ \mathcal{L}(t',t)$ represents the linear time
evolution map of the master equation \eqref{me} from time $t$ to $t'$. %The autocorrelation function can be decomposed in orthonormal output modes $v_i$ as
%\begin{align}\label{g11}
%    \mathcal{G}^{(1)}(t_1,t_2) = \sum_i n_i v^*_i(t_1)v_i(t_2),
%\end{align}
%where $n_i$ is the mean occupation number of the $i$'th mode. The eigenmodes $v_i(t)$ illustrate the shape of each mode in time. 
The time-dependent spectrum related to the autocorrelation function in Eq. \eqref{g1} is found by the Fourier transform %\cite{timefrequencyChapter} 
\begin{align}\label{Fwt}
    \mathcal{I}(\omega,t) = \int_{-\infty}^{+\infty} \mathrm{d}s \,\,\,\,\mathcal{G}^{(1)}\bigg(t+\frac{s}{2},t-\frac{s}{2}\bigg)\,\, e^{-i\omega s},
\end{align}
which provides information about the time-dependent frequency content of the output field.

Output field spectra for a nonlinear resonator and a linear resonator are shown in Fig. \ref{Fig1}, where in both cases the resonator is initialized in the Fock state $\ket{\psi} = \ket{5}$. For the parameters in the figure caption, the time-dependent spectrums show the emission of a wide range of frequencies for both linear and non-linear resonators. The spectrum in Fig. \ref{Fig1} (a), shows a visible gap between the frequency pertaining to each emitted photon, proportional to the amount of the Kerr nonlinearity $2\chi$. The insert shows that multiple output field modes are populated in the field emitted by the nonlinear resonator. 

For comparison, in Fig. \ref{Fig1}(b), we assume a linear resonator, $\chi=0$, with a time-dependent frequency $\omega(t)$ chosen to give a similar frequency range of the emitted field as in (a). The insert in panel (b) shows that the output field in this case only occupies a single (chirped) mode. 
%This clearly illustrates that the non-linearity entangles the Fock state and spatio-temporal components and yields a multimode output field, while a linear resonator, even with a frequency chirp, retains a single mode. 
In the next section, we model a realistic quantum emitter implemented in superconducting circuits and we assess how the non-linear elements used for tuning the emission process affect the mode character and purity of the emitted quantum state.

\section{Physical Model}\label{SLR} 
\begin{figure}[ht!]
\centering
\includegraphics[width=.5\textwidth]{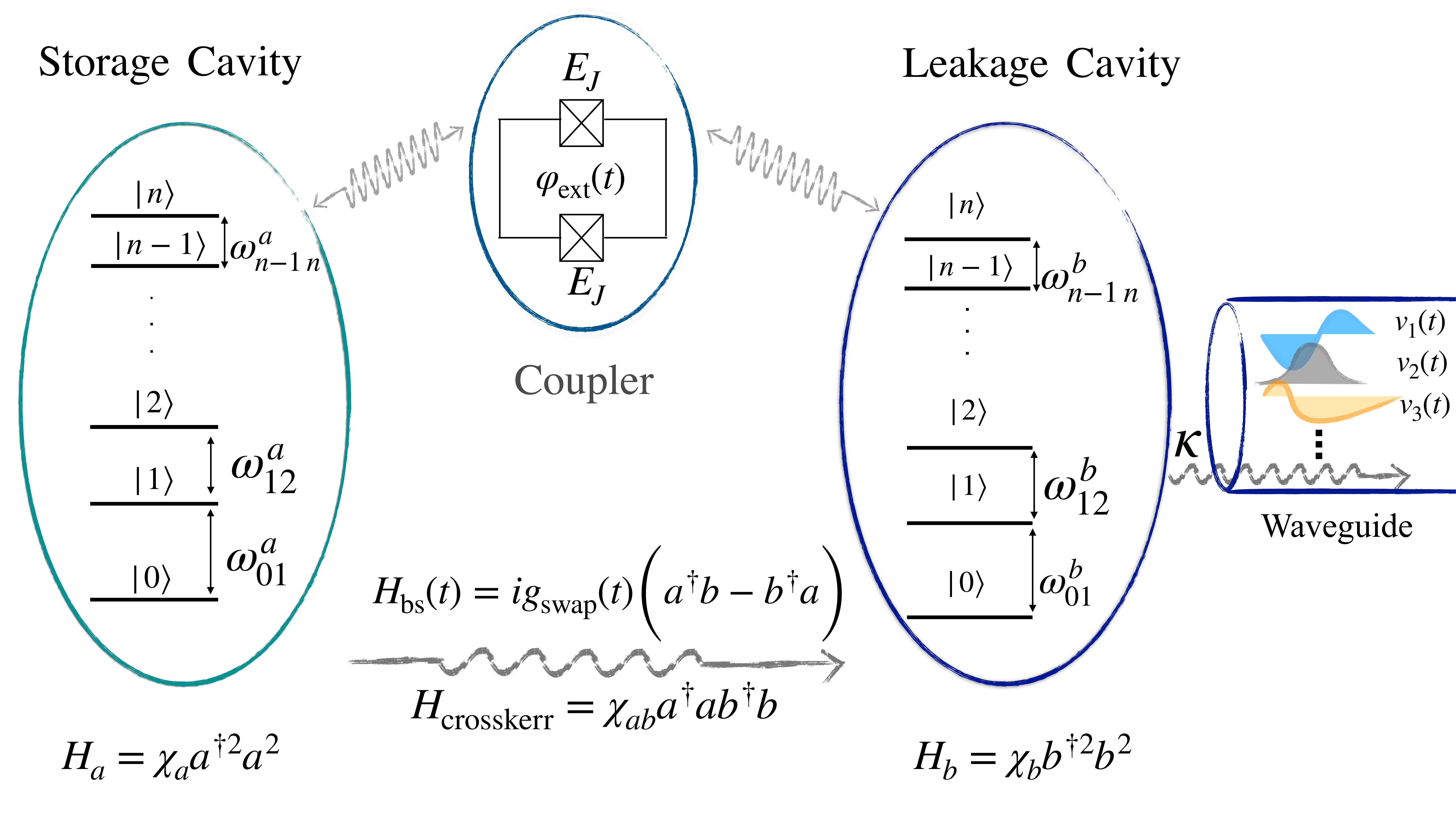}
\caption{Schematic of the emitter. The non-linear coupler interacts with both the storage and the leakage cavity. By driving the coupler at the frequency $\omega^a_{01} - \omega^b_{01}$, it implements a frequency conversion and resonant transfer of quanta between the storage and leakage cavity, as governed by a beam splitter interaction $H_{bs}(t)$ between the two cavities at frequencies $\omega^a_{01}$ and $\omega^b_{01}$. Due to the nonlinearity of the coupler, the beam splitter  interaction is accompanied by self-Kerr and cross-Kerr non-linear interactions; see Eq. \eqref{H1}, which leads to an output field populating several mode´s $v_1(t),v_2(t),v_3(t),...$ .}
\label{F2-cavitycoupler}
\end{figure}

Several approaches have been used to map a stationary resonator mode to a propagating pulse mode.
To optimally control the release of a quantum state, we consider the storage system (cavity), initiated in the desired quantum state $|\psi\rangle$, described by the creation and annihilation operators $a^\dagger,a$ and frequency $\omega_a$.
The storage cavity is dispersively coupled to a flux-tunable transmon \cite{tinkham2004introduction, PhysRevA.76.042319,blais2021circuit,rasmussen2021superconducting} described by the operators $c^\dagger,c$ and frequency $\omega_c$, through the coupling strength $g_{ac}\ll(\omega_c-\omega_a)$. In addition to the storage cavity, the coupler is also dispersively coupled to a leakage cavity, described by frequency $\omega_b$ and operators $b^\dagger,b$, with the coupling $g_{bc}\ll(\omega_b-\omega_c)$;  see Fig. \ref{F2-cavitycoupler}. 

The entire system is described by the effective Hamiltonian 
\begin{align}\label{HH}
    H_S(t) = &\omega_a a^\dagger a+\omega_b b^\dagger b+\omega_c c^\dagger c \nonumber\\
&+2E_{J}\bigg[\cos(\frac{\varphi_{\mathrm{ext}}(t)}{2})-\cos(\frac{\varphi_{\mathrm{dc}}}{2})\bigg] \frac{\varphi^2_c}{2} \nonumber\\
  &-2E_{J}\cos(\frac{\varphi_{\mathrm{ext}}(t)}{2})\frac{\varphi^4_c}{4!},
\end{align}
where $\varphi_c$ and $E_J$ correspond to the reduced flux operator of the transmon and the energy of the junction, respectively. The AC flux drive on the coupler is $\varphi_{\mathrm{ext}}(t) = \varphi_{\mathrm{dc}} + F(t)\sin(\omega_d t)$  which is described by the time-dependent amplitude $F(t)\equiv \delta \tanh{(t/t_0)}\ll 1$ and frequency $\omega_d$, where $\delta$ and $t_0$ correspond to the amplitude and rate of turning on the drive, respectively. For more details on the derivation of the Hamiltonian \eqref{HH} see Appendix \ref{SetupQEC}.

In the dispersive regime, the reduced flux of the coupler is found as the superposition of all dressed modes \cite{nigg2012black,marcos2013superconducting} 
\begin{align}\label{dressed2}
    \varphi_c =& \frac{2\pi}{\phi_0}\phi_c,\nonumber\\
    \phi_c =& \frac{\lambda_a (a+a^\dagger)+\lambda_b (b+b^\dagger)+\lambda_c (c+c^\dagger)}{\sqrt{2}}  \equiv\frac{A+A^\dagger}{\sqrt{2}},
\end{align}
where $\phi_c$ is the flux of the coupler and the coefficients $\lambda_{a,b,c}$ are described in  Eq. \eqref{dressed1}. 
Using the Taylor expansion \eqref{potential1}, we insert Eq. \eqref{dressed2} in the Hamiltonian \eqref{HH}, 
thus the second line of Eq. \eqref{HH} is obtained as 
\begin{align}\label{L1}
    H^l(t) = 
    \frac{-2E_J\pi^2}{\phi_0^2}\Bigg[&\frac{\sin(\frac{\varphi_{\mathrm{dc}}}{2}) F(t)\sin(\omega_d t)}{2} \nonumber\\ &+ \frac{\cos(\frac{\varphi_{\mathrm{dc}}}{2})}{8}F(t)^2\sin(\omega_d t)^{2}\Bigg](2A^\dagger A + \bar{\lambda}),
\end{align}
where $\bar{\lambda} = [A,A^\dagger] $. The third line of Eq. \eqref{HH} (fourth order of the flux operator),  provides the nonlinear interactions as follows
\begin{align}
  %H_{nl} \approx & \frac{-\pi^4E_j\cos(\frac{\varphi_{\mathrm{ext}}(t)}{2})}{3\phi_0^4}(A+A^\dagger)^4, \nonumber\\
   H_{nl} \approx &\frac{-\pi^4E_J\cos(\frac{\varphi_{\mathrm{ext}}(t)}{2})}{3\phi_0^4}\Bigg[12 \bar{\lambda} A^\dagger A +6 A^\dagger A^\dagger A A \Bigg],
\end{align}
where we just keep the terms conserving the energy.

\begin{figure}[ht!]
\includegraphics[width=.48\textwidth]{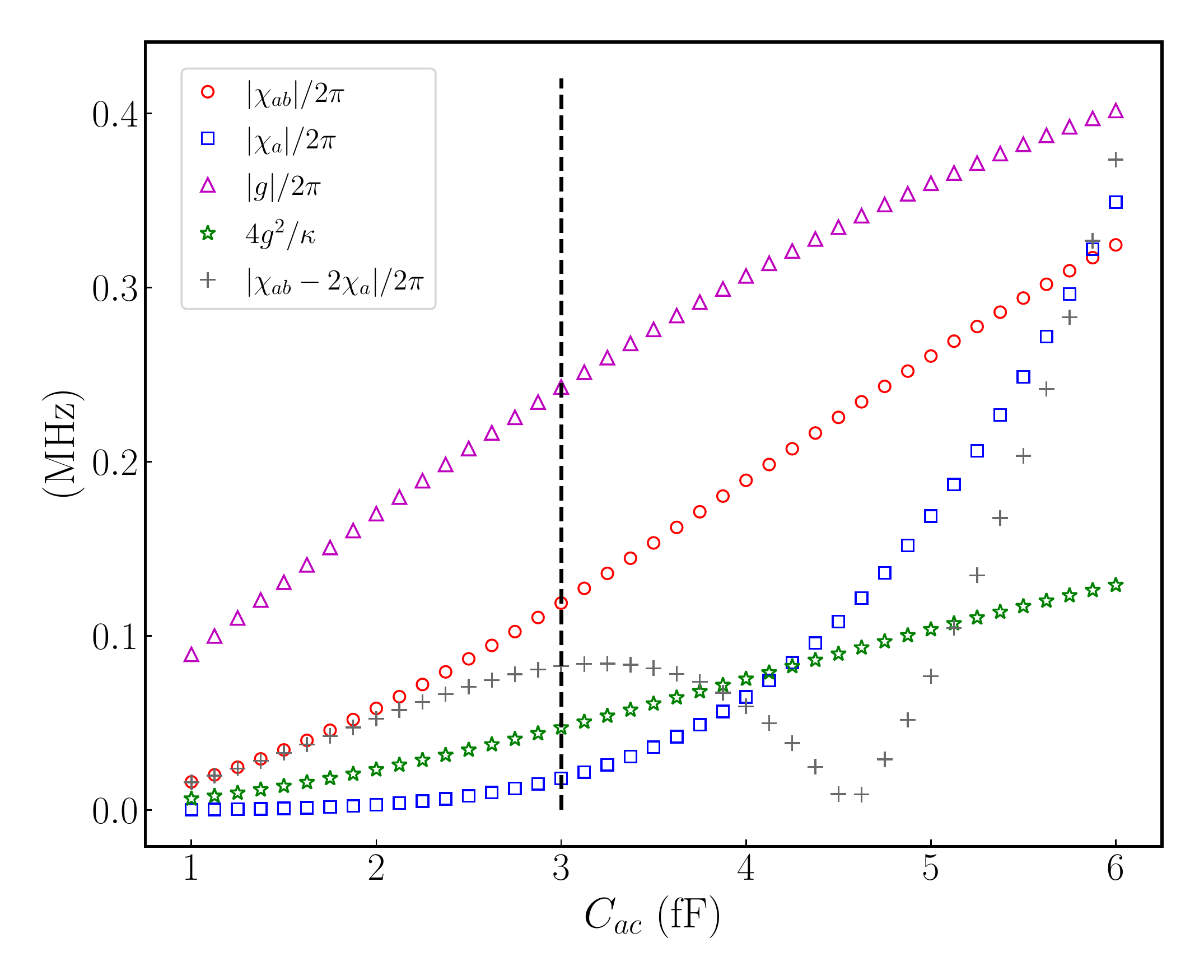}
\caption{ The change of the nonlinear parameters $\chi_a,\chi_{ab}$, swap rate $g$, detuning from resonant interaction $\chi_{ab}-2\chi_a$, and Purcell rate $4g^2/\kappa$ as functions of the value of the coupling capacitance $C_{ac}$, where $1/\kappa = 0.2 \, \mu s$. The vertical black dashed line corresponds to the parameters used to simulate the lower panels in Fig. \ref{Fig4} and Fig. \ref{wigner}.} 
\label{Nparameter}
\end{figure}

We assume that the coupler is initiated in the ground state and mediates the resonant frequency conversion without itself being excited. This permits the elimination of its quantum degrees of freedom at all times.
If we consider $\omega_d = \omega_b - \omega_a$ and utilize the rotating wave approximation and transform to the rotating frame interaction picture with respect to $\omega_a a^\dagger a+\omega_b b^\dagger b+\omega_c c^\dagger c$, the Hamiltonian is obtained as 
\begin{align}\label{H1}
    H_S(t) = &S_a (t) a^\dagger a+S_b(t) b^\dagger b+\chi_{a}a^{2\dagger}a^2+\chi_{b}b^{2\dagger}b^2\nonumber\\
    &+\chi_{ab}a^{\dagger}a b^{\dagger}b - i g_{\mathrm{swap}}(t)(b^{\dagger}a-a^{\dagger}b),
\end{align}
where the parameters are as follows
\begin{align}\label{Coeff5}
    \chi_{a(b)} =& \frac{-2\pi^4E_J\cos(\frac{\varphi_{\mathrm{dc}}}{2})}{\phi_0^4} \lambda_{a(b)}^4, \nonumber\\
    \chi_{ab} =& \frac{-8\pi^4E_J\cos(\frac{\varphi_{\mathrm{dc}}}{2})}{\phi_0^4} \lambda_a^2\lambda_b^2,\nonumber\\
    g_{\mathrm{swap}}(t) =& (\frac{\pi^2}{\phi_0^2}-\frac{\bar{\lambda}\pi^4}{\phi_0^4})E_J  \sin(\frac{\varphi_{dc}}{2}) \lambda_a \lambda_b F(t),\nonumber\\
    S_{a(b)}(t) =& (\frac{\bar{\lambda}\pi^4}{\phi_0^4}-\frac{\pi^2}{\phi_0^2})\frac{E_J \cos(\frac{\varphi_{\mathrm{dc}}}{2})\lambda_{a(b)}^2}{4}F(t)^2,
\end{align}
and $S_{a(b)}(t) $ are the Stark shifts induced by the flux drive. The dressed mode coefficients $\lambda _{a(b)}$ in Eq. (\ref{dressed2},\ref{Coeff5}) are proportional to the ratio $\lambda _{a(b)} \propto g_{a(b),c}/\Delta_{a(b),c}$, where $g_{a(b),c} $ and $\Delta_{a(b),c}$ correspond to the coupling strength and detuning between the coupler and the storage (leakage) cavity. The coupler is coupled to the storage (leakage) cavities through the coupling capacitance $C_{ac}(C_{bc})$ where by changing the capacitance strengths, different values of the conversion rates $g_{\mathrm{swap}}$ and also nonlinear terms $\chi_{ab},\chi_{a(b)}$ will be obtained; see Fig. \ref{Nparameter}. As shown in Fig. \ref{Nparameter}, the reduction (increase) of the non-linearities $\chi_{a(b)} \propto \lambda_{a(b)}^4$ yields a similar effect on the swap rate $g_{\mathrm{swap}}(t)\propto \lambda_{a}\lambda_{b}$, as mentioned $\lambda_{a(b)}\propto g_{a(b),c}\propto C_{a(b)c}$.  It is worth noting that while a stronger amplitude of the drive $F(t)$, accelerates the transfer and release process, in this regime the Hamiltonian of the system acquires higher-order nonlinear interactions. 
%In addition, in the experiment, there is always a threshold on the strength of the drive amplitude.

 The leakage cavity decays to the waveguide with a constant decay rate $\kappa$.
The analysis of the emitted radiation is equivalent to the one presented for the toy model in Sec. \ref{Theory}, with the master equation,
\begin{align}\label{me2}
    \frac{\partial \varrho(t)}{\partial t} = -i[H_S(t), \varrho(t)] + \kappa\Bigg(b \varrho(t)b^\dagger - \frac{1}{2}[b^\dagger b, \varrho(t)]\Bigg),
\end{align}
where $H_S(t)$ is given in Eq.
\eqref{H1},
and the Lindblad operator $\sqrt{\kappa} b$ describes the dissipation of the leakage cavity to the waveguide. As for the toy model in Eq. \eqref{g1}, the mode decomposition of the first-order correlation function of the field operator $b$, determines the most populated orthonormal output modes.

\begin{figure*}[ht!]
\centering
\includegraphics[width=1\textwidth]{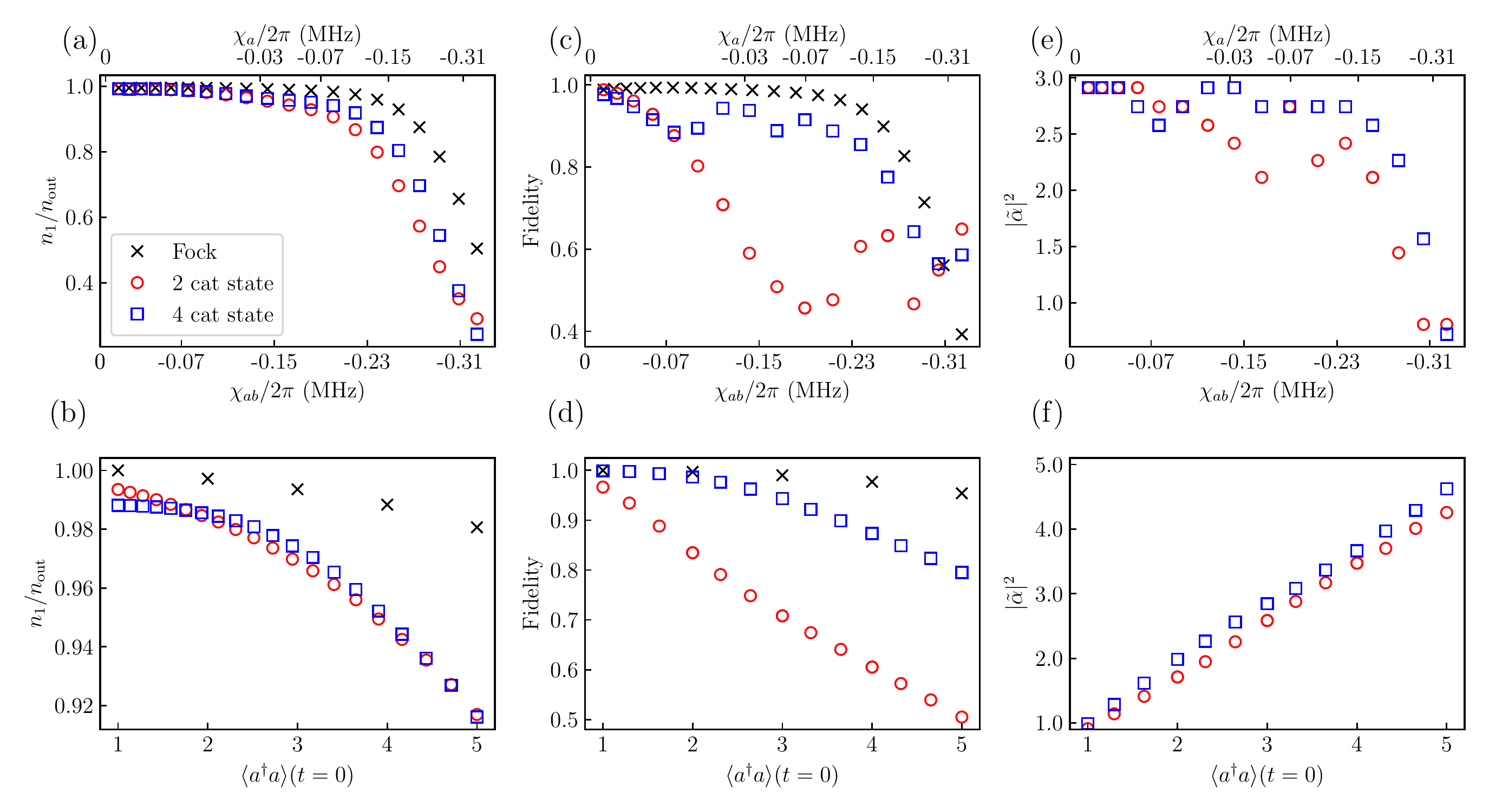}
\caption{The characteristics of the output field for different initial quantum states such as a FS $\ket{n}$, a TCCS \eqref{2cat}, and a FCCS \eqref{4cat}. The upper panels show (a) the relative occupation of the most populated mode, (c) the state fidelity of the state occupying that mode with respect to the optimal target state, see Eq. \eqref{Fidelity}, and (e) the value of the cat state amplitude $\Tilde{\alpha}$ that has the maximum fidelity with the state  occupying the most populated mode.The results are shown for the initial FS $\ket{n=3}$ and FCCS and TCCS with amplitude parameter $\alpha = \sqrt{3}$. They are calculated as a function of the non-linear coupling strengths $\chi_{ab}$ and $\chi_{a}$ indicated along the lower and upper axes, which in turn correspond to a varying capacitance between the storage cavity and the coupler in the physical system (see Fig. \ref{Nparameter}). The self-Kerr $\chi_{b}$ is not specified as the leakage cavity is mostly occupied by the vacuum and the occasional one-photon state.  In the lower panels, we assume non-linear interaction parameters $\chi_{ab}/2\pi= \SI{-0.11}{MHz}$, $\chi_a/2\pi= \SI{-0.017}{MHz}$, and $\chi_b/2\pi= \SI{-0.04}{MHz}$, and we show the same quantities as in the upper panels but for different values of the initial mean photon number $\langle a^\dagger a\rangle(t=0)$; see Sec. \ref{MOF} for details.}
\label{Fig4}
\end{figure*}
 
As we saw in Sec. \ref{Theory}, non-linear terms in the Hamiltonian cause the output field to populate several temporal field modes of the waveguide. Since the self Kerr and cross Kerr terms $(\chi_{a}a^{2\dagger}a^2, \chi_{b}b^{2\dagger}b^2$ and  $\chi_{ab}a^{\dagger}b^{\dagger}ab)$ are inevitable consequences of the tunable coupling in our system, the output field will indeed populate many temporal modes, as will be seen in the next section.

In the limit where $\kappa$ is much larger than the other couplings in the master equation, it is possible to adiabatically  eliminate the leakage cavity and obtain an effective Markovian master equation for the storage cavity mode  with a Purcell damping rate $\simeq 4g_{\mathrm{swap}}^2/\kappa$; green dots in Fig \ref{Nparameter}. While we do not rely on the quantitative validity of this effective treatment in our numerical studies, we will refer to the value of the Purcell rate in the analysis of the results. 

Here, we also note that due to the presence of other dissipation channels, 
%with rates $\kappa_i$, 
the release process cannot be made arbitrarily slow. Hence, in the experiment, there will be a trade-off between the loss to many less populated modes in the rapid-release regime and to other dissipation channels in the slow-release regime.
%the time of the state transfer between the cavities $t_{\mathrm{swap}}\propto 1/g_{\mathrm{swap}}$, the effective loss rate $4g_{\mathrm{swap}}^2/\kappa$, mediated by the leakage cavity, and the effect of the Kerr non-linearities $\chi_{ab},\chi_a$. 

%\textcolor{blue}{I would suggest moving figure 3 to a subfigure in figure 4, I think i takes up unnessecarily much space in its current form and i think it would be much nicer to have the information of figure 3 in the vicinity when looking at figure 4.}\change{MARYAM. My suggestion is to put it separately, as it has many parameters with different colors and shapes. Fig 4 is also crowded enough. also, it brings a lot of changes and references to figures in different parts of the paper. ;) }

%%%%%%%%%%%%%%%%%%%%%%%%%%%%%%%%%%%%%%%
%%%%%%%%%%%%%%%%%%%%%%%%%%%%%%%%%%%%%%%
%%%%%%%%%%%%%%%%%%%%%%%%%%%%%%%%%%%%%%%
%%%%%%%%%% RESULT SECTION %%%%%%%%%%%%%
%%%%%%%%%%%%%%%%%%%%%%%%%%%%%%%%%%%%%%%
%%%%%%%%%%%%%%%%%%%%%%%%%%%%%%%%%%%%%%%
%%%%%%%%%%%%%%%%%%%%%%%%%%%%%%%%%%%%%%%

\section{Release of Fock states and cat states}\label{sec:results}

As illustrated in Sec. \ref{Theory}, a linear resonator emits any initial quantum state into a single spatiotemporal mode determined by the time-dependent outcoupling strength. However, the temporal shape of the output field from a  nonlinear emitter is correlated with the photon number contents. 
As discussed in the preceding section, the strength of the coherent swap coupling between the cavities, the effective Purcell decay rate of the storage cavity, and the non-linearity are correlated and vary with the physical parameters of the circuit. In this section, we characterize the output field from the resonator using the variation of the parameters corresponding to the coupling capacitance strength between the storage cavity and the coupler which are shown in \ref{Nparameter}.   
%%Hence we consider different values of the coupling strength $g_{ac}\propto C_{ac}$ (without violating the dispersive regime condition $g_{a(b)c}\ll \Delta_{a(b)c}$), where $C_{ac}$ is the capacitance that couples the storage cavity to the coupler; see Fig. \ref{F1-cavitycoupler}. With a fixed maximum value $\delta \ll 1$ for the drive amplitude $F(t) = \delta \tanh(t/t_0)$, the change of the swap strength with respect to different values of the nonlinear interaction is shown as the purple y-axis in the panel \ref{F3} (e), where we consider the drive is in the maximum value $F(t) = \delta$. This figure clearly shows the trade-off between the time of the transformation $t_{\mathrm{swap}}\propto 1/g_{\mathrm{swap}}$ and the effect of the nonlinear interactions. In the following, we investigate the effect of these correlated parameters $\{g_{\mathrm{swap}},\chi_{a},\chi_{ab},\chi_{b}\}$ on different initial states and photon numbers.

In Fig. \ref{Fig4}, we investigate the output field of the emitter for different quantum states such as a Fock state (FS) $\ket{\psi} = \ket{n}$, a two-component cat state (TCCS) 
\begin{align}\label{2cat}
    \ket{\psi} \propto  \ket{\alpha}+\ket{-\alpha} \propto \sum_{n=0}^\infty \frac{\alpha^{2n}}{\sqrt{(2n)!}}\ket{2n},
\end{align}
which is a promising candidate for correcting dephasing errors \cite{PhysRevA.59.2631,PhysRevLett.100.030503} , and a four-component cat state (FCCS)
\begin{align}\label{4cat}
    \ket{\psi}\propto \ket{\alpha}+\ket{-\alpha}+\ket{i\alpha}+\ket{-i\alpha}
    \propto \sum_{n=0}^\infty \frac{\alpha^{4n}}{\sqrt{(4n)!}}\ket{4n},
\end{align}
which can be used for quantum storage and communication in the  presence of photon loss \cite{PhysRevLett.111.120501,mirrahimi2014dynamically}.

Fig. \ref{Fig4} (a)  shows the relative population of the most populated mode $n_1/n_{out}$ as a function of the Kerr nonlinearities. The ratio $n_1/n_\mathrm{out}$ reveals the multimode character of the output field released from a three-photon FS and TCCS and FCCS, composed of coherent states with amplitude $\alpha^2 = 3$.  As one expects, the output field becomes more multimode with higher values of nonlinearity and with higher photon numbers. We observe that the FS and the FCCS yield a higher single-mode content than the TCCS with the same photon number. This difference arises partly because the FS and FCCS populate a single, dominant Fock state component, while the TCCS populates both $\ket{n=2}$ and $\ket{n=4}$ components, which are dephased by the Kerr effect and couple to different frequency components of the output field.  Panel \ref{Fig4} (b) shows $n_1/n_\mathrm{out}$ as a function of the initial mean photon number of the storage cavity for the same states using fixed nonlinear parameters, corresponding to the dashed vertical line in Fig. \ref{Nparameter}. As expected, with a higher number of photons, the resonator non-linearity leads to an output field occupying more modes.

In Fig.\ref{Fig4}, panel  (a), the output field is almost single mode, $n_1/n_{\mathrm{out}} \simeq 1$, until  $\chi_{ab}/2\pi$ exceeds the value  $-0.23$ MHz or equivalently the value $C_{ac} \simeq 4.5 $fF in Fig \ref{Nparameter}. This can be ascribed to the release of photons being meditated by the process, $\ket{n,0}_{a,b}\xrightarrow{\text{swap}}\ket{n-1,1}_{a,b}\xrightarrow[]{\kappa}\ket{n-1,0}_{a,b}$. The energy difference between the states $\ket{n,0}_{a,b},\ket{n-1,1}_{a,b}$ is $\Delta E_n = (n-1)(2\chi_a-\chi_{ab})$ where a small additional shift is omitted, see Eq. \eqref{H1}. If $\chi_{ab} \simeq 2\chi_a$, the transfer is resonant and faster which results in a more single-mode character. According to Fig. \ref{Nparameter}, the energy difference has a smooth behavior until $C_{ac} \simeq 4.5$fF and hereafter, the transition states rapidly become non-resonant and the slow release in combination with the  nonlinearity causes the increasing multimode character, witnessed by the reduction of the population $n_1/n_{\mathrm{out}}$ and similar reductions in the other quantifiers of the output field.

\section{Characterizing the most populated mode}\label{MOF}%What is the quantum state occupying the most populated mode?}\label{MOF}

\begin{figure}[ht!]
\includegraphics[width=.52\textwidth]{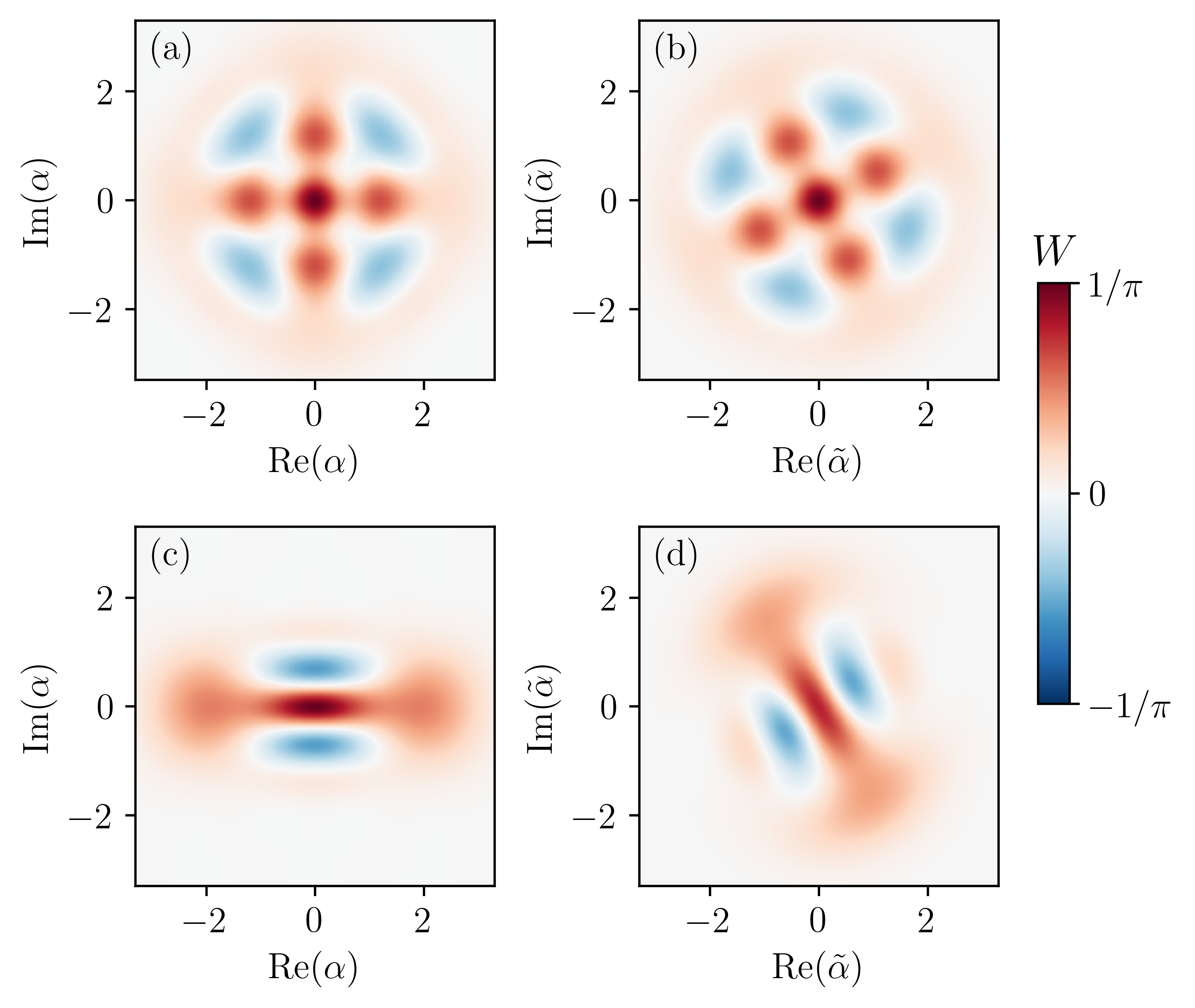}
\caption{Wigner function of the initial FCCS (a) and the state of the most populated mode of emission (b), the initial TCCS (c) and the corresponding highest fidelity single mode output state (d). The amplitude of the initial cat states is $|\alpha^2| = 2$ and the nonlinear interactions are $\chi_{ab}/2\pi= -\SI{0.11}{MHz} , \chi_a/2\pi= -\SI{0.017}{MHz} $, and $\chi_b/2\pi= \SI{-0.04}{MHz}$; the same as panels \ref{Fig4} (b,d,f) . The amplitude of the single mode output cat state for panel (b) is $|\Tilde{\alpha}|^2 =1.96 $ with a corresponding fidelity $\mathcal{F}_{\mathrm{4cat}} = .97$. Panel (d) shows the emitted single mode TCCS obtained with the optimal drive rate $t_0$ for the flux drive $F(t) = \delta \tanh(t/t_0)$. This state has a lower amplitude $|\Tilde{\alpha}|^2 =1.66$ and the fidelity $\mathcal{F}_{\mathrm{2cat}} = .85$; See Sec. \ref{SEC:VA} for more explanation.} 
\label{wigner}
\end{figure}

In this section, we address the quantum state contents of the most occupied mode $v_1$. This is done by the 
formalism introduced in \cite{kiilerich2019input}, which, for the theoretical calculation assumes a downstream ideal linear cavity with mode operators $d,d^\dagger$, coupled to the system by the interaction Hamiltonian \begin{align}\label{Hbv}
    H_{b,v_1}(t) = \frac{i\sqrt{\kappa} }{2}[g_{v_1}(t)d^\dagger b -g^*_{v_1}(t)b^\dagger d].
\end{align} 
The time-dependent coupling between the cavity and waveguide
\begin{align}\label{gv}
    g_{v}(t) = -\frac{v_1^*(t)}{\sqrt{\int_0^t \mathrm{d}t' |v_1(t')|^2}},
\end{align}
ensures that the cavity captures the contents of the temporal mode $v_1(t)$. The dynamics of the cascaded system is described by the Lindblad master equation with total Hamiltonian $H_T = H_S(t)+H_{b,v_1}(t)$ and a single  Lindblad operator describes dissipation to the waveguide, $L_0(t) = \sqrt{\kappa} b + g_{v_1}^*(t)d$, representing the interference between the emitted field of the leakage cavity and the ideal downstream cavity. This form of the master equation ensures the cascaded nature of the propagation of the fields, \cite{carmichael1993quantum,gardiner1993driving}.   

In the three oscillator description (a,b,d), the ideal state transformation is $\ket{\psi,0,0}_{a,b,d}\rightarrow \ket{0,0,\psi}_{a,b,d}$. While the $a$ and $b$ oscillators may, indeed, be emptied with certainty, cavity $d$, will, in general, be occupied by a mixed state $\varrho_d$, as it is correlated with other modes of the output field. %The fidelity of the state transformation up to a linear rotation phase is 
%where $T$ is the temporal extension of mode $v_1(t) $.

Because the output field is not a single mode, the number of photons in the most populated output mode $v_1$ is less than the initial number of photons in the storage cavity. As shown in Figs. \ref{wigner} (b,d),  the output state in that mode  may still be a cat-like state $\ket{\tilde{\psi}}$, however, with a modified amplitude $\tilde{\alpha}$. We thus vary the parameter $\tilde{\alpha}$ in order to maximize the fidelity 
\begin{align}\label{Fidelity}
    \mathcal{F} = \bra{\tilde{\psi}} \varrho_{\mathrm{d}}(T)\ket{\tilde{\psi}},
\end{align}
over TCCS and FCCS cat states $\ket{\tilde{\psi}}$ 
where $T$ is a time well after the emission of the temporal mode $v_1(t)$.
For the Fock state fidelities, we calculate the population of the same Fock state  $\ket{\tilde{\psi}}=\ket{\psi}$ as the initial state of the storage cavity.

Panel \ref{Fig4} (c) shows the fidelity $\mathcal{F}$ of the quantum state occupying the most populated mode $v_1$ as a function of the Kerr nonlinearities. For all three initial states, the fidelity is higher in the low Kerr regime, however, the fidelity of TCCS is affected more than the other quantum states which we ascribe to the TCCS occupying more Fock components. 
In panel \ref{Fig4} (c), in addition to the rapid reduction of the fidelity, a revival of the fidelity of the TCCS is observed in the higher nonlinear regime. This can be explained by considering the different phases $\theta_2$ and $\theta_4$ induced by the self Kerr and cross-Kerr coefficients $\chi_a,\chi_{ab}$ on the Fock components $\ket{2}$ and $\ket{4}$, respectively. For a certain amount of the nonlinearities, the phases can be related by $\theta_4 \approx \theta_2+2\pi k$, recovering fidelity of the TCCS; see Fig. \ref{SC} in appendix \ref{SC1} for visualizations of the Wigner functions corresponding to Fig. \ref{Fig4} (c). 

Panels \ref{Fig4} (e) and (f) show the size of the cat states $\tilde{\alpha}$ of the mode $v_1$ as a function of the nonlinearities and the initial photon number, respectively. In Fig. \ref{Fig4} (e), for both cat states, the amplitude $\tilde{\alpha}$ decreases with increasing the nonlinearity as the less number of photons populates the most populated mode; see panel \ref{Fig4} (a). Panel \ref{Fig4} (d) shows the fidelity of the FS, TCCS, and FCCS as the function of the initial mean photon number where the higher photon numbers, as expected, yield more reduction in the fidelity.

In the low photon number regime, $n<2$, the loss of photons to other modes is noticeable, Fig. \ref{Fig4} (b), but is not dominating over the reduction of the fidelity coming from the Kerr rotation. In the recent experiment \cite{axline2018demand}, transferring TCCS and FCCS between a sender and a receiver, the dynamics is determined by a similar Hamiltonian as Eq. \eqref{H1}. In the supplementary material of \cite{axline2018demand}, the fidelity reduction due to an effective Kerr rotation combined with photon loss is discussed. Our formalism gives a very similar picture, where we can quantify the cavity dephasing due to the population of multiple output modes. We find a one percent reduction of the population of the most populated output mode, which agrees with the analysis in \cite{axline2018demand}.

\subsection{Catching the optimal cat states} \label{SEC:VA}

%As shown in the panel \ref{Fig4} (b), the number of photons in the output mode $v_1$ is less than the total number of photons, however, the output state of that mode  may still be a cat-like state, but with a lower amplitude parameter $\tilde{\alpha}$. In Fig. \ref{Fig4}, panels (e,f), we determine the TCCS and FCCS cat states with amplitude $\Tilde{\alpha}$ which has the highest overlap with the quantum state contents of the  $v_1$ mode. We see that the amplitude of the optimal reabsorbed cat state decreases with increasing non-linearity. 
%%%%%%%%%

As discussed in the previous section, the Kerr rotations affect the FCCS less, as only two Fock components $\ket{0},\ket{4}$ are essentially populated and the effect can to a large extent be described by an effective phase acquired by $\ket{4}$. One realization of the initial and the released FCCS is shown in Fig. \ref{wigner}, panels (a) and (b), respectively. The initial amplitude is $|\alpha^2| =2$ and the emitted single-mode state has the highest fidelity with a FCCS with amplitude $|\tilde{\alpha}^2| = 1.96$; the same as panel \ref{Fig4} (f) and with the fidelity of $\mathcal{F}_{\mathrm{4cat}} =.97$ which is shown in \ref{Fig4} (d).

In contrast to FCCS, the TCCS has three main Fock components $\ket{0},\ket{2}$ and $\ket{4}$, thus involving two different effective phases, which in general can not be modeled by a single rotation of the Wigner function. To improve the fidelity for a specific set of nonlinear parameters, we optimize the flux drive on the coupler to minimize the effect of the nonlinear rotations and find an approximate TCCS with relation $\theta_4 \approx \theta_2 + 2\pi k$ and amplitude $\tilde{\alpha}<\alpha$. Panel (c) shows an initial TCCS with $|\alpha|^2 =2$, and panel (d) shows the state released in the most occupied mode using the optimal drive rate $ t_0 = 7.3 \SI{}{\mu s}$, which is almost 5 times slower than the one employed for Fig. \ref{Fig4} and Fig. \ref{wigner} (a,b). The amplitude of the reabsorbed TCCS is $|\Tilde{\alpha}|^2 = 1.66$ with fidelity $\mathcal{F}_{\mathrm{2cat}} =.85$ up to the  arbitrary linear rotation illustrated in Fig. \ref{wigner} (d).

\section{Summary}\label{Sum}
%\change{We need to compare the result to the Yale setup. perhaps here. }
We studied the characteristics of the quantum state released from a realistic nonlinear emitter. To optimally release the desired quantum states into a propagating mode, we utilized a flux-tunable coupler to transfer the quantum state from the storage cavity to the waveguide. We have shown that due to the nonlinear interactions in the emitter, the output field obtains a multimode character, where the shape of the modes and their photon population become correlated.  We investigated the output field for Fock states and two- and four-component cat states. We also studied the adjustment of the flux drive to emit an optimal cat state into the most populated mode. Our results showed that in the low photon number regime, the fidelity reduction due to the nonlinear interactions is clearly noticeable (on the percent level), but it is not the dominant contribution to the experimentally observed reduction in fidelity, seen in recent experiments
\cite{pfaff2017controlled,axline2018demand}. 

Our calculation and simulation results illustrate a trade-off between the speed of emission and the effective nonlinearities of the emitter. This may suggest that using more elaborate couplers with more tunability, e.g a SNAIL-based coupler \cite{frattini20173}, may improve the fidelity of the beam-splitter gate $\propto H_{bs}(t)$ \cite{chapman2022high}. 
However, according to our formalism, in aiming for high-fidelity state transfer of multiphoton states, the multimode aspects of the transmitted field are unavoidable and need to be taken into account.

Lastly, it is worth to note that, to actually catch  a single mode using a linear receiver, it is enough to use the time-inversed drive compared to the one used in a linear transmitter \cite{cirac1997quantum}. A realistic non-linear emitter, however, generates a multi-mode output field, and we would need a non-linear receiver to optimally reabsorb the output field. The concept of time reversal can be used as a guiding principle, but how to find such a receiver in practice is still an open question.

\begin{acknowledgements}
M. Khanahmadi and G. Johansson acknowledge Simone Gasparinetti for useful comments and the support from Knut and Alice Wallenberg Foundation through the Wallenberg Center for Quantum Technology (WACQT). M. M. Lund and K. M\o lmer acknowledge support from Carlsberg Foundation
through the “Semper Ardens” Research Project QCooL. 
\end{acknowledgements}
\appendix
\section{Lagrangian of the quantum circuit of the Emitter}\label{SetupQEC}
\begin{figure}[ht!]
\centering
\includegraphics[width=.48\textwidth]{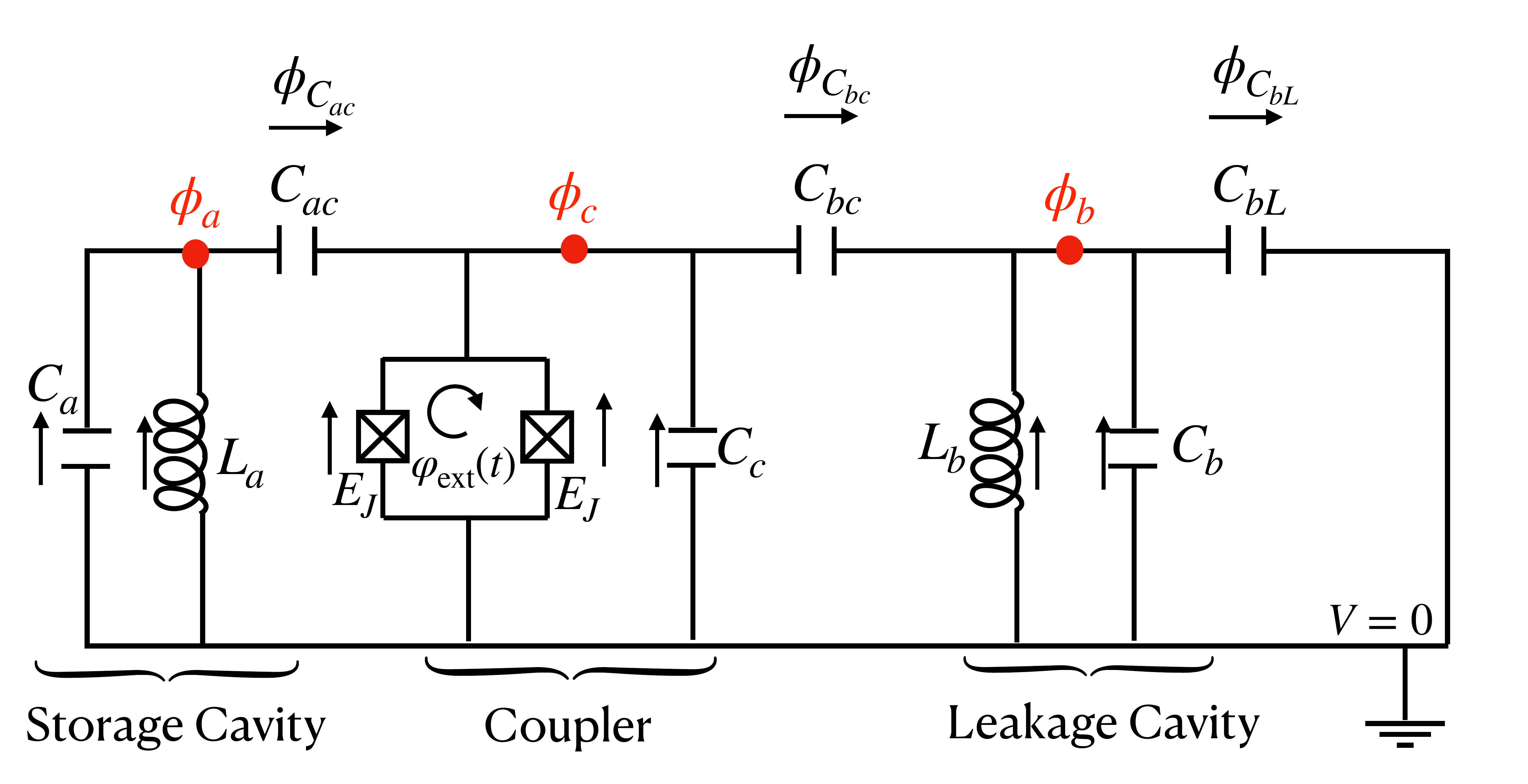}
\caption{ Quantum circuit of the emitter. The coupler is derived to properly transfer a quantum state from storage to the leakage cavity. Both cavities are shown by lumped elements which capacitively are coupled to a symmetry flux tunable transmon.}
\label{F1-cavitycoupler}
\end{figure}
The quantum circuit of Fig. \ref{F2-cavitycoupler} is shown in detail in Fig. \ref{F1-cavitycoupler}. We consider the two cavities as lumped elements including capacitances $C_{a(b)}$ in parallel with the inductances $L_{a(b)}$. The coupler is described by the total capacitance $C_c$, junction energy $E_j$, and the AC flux drive $\varphi_{\mathrm{ext}}(t)$. The coupler capacitively is coupled to the storage and leakage cavity through the capacitance $C_{ac},C_{bc}$, respectively. The capacitance $C_{bL}$ is the coupling between the leakage cavity and the transmission line which gives rise to a minor frequency shift of the leakage cavity. The red dots show the flux of each system with which the Lagrangian and the Hamiltonian of the system can be evaluated.

To properly drive the effective Hamiltonian we first evaluate the lagrangian of the quantum circuit 
$ \mathcal{L} = \mathcal{T}-\mathcal{V}$ where $\mathcal{T}$ and $\mathcal{V}$ correspond to the kinetic and potential energy, respectively.

According to Fig. \ref{F1-cavitycoupler}, the kinetic energy is evaluated as
\begin{align}\label{T}
    \mathcal{T} = \frac{1}{2}\Bigg[& C_{a} \Dot{\phi}_{a}^2 + C_{ac} \Dot{\phi}_{C_{ac}}^2 +C_{c} \Dot{\phi}_{c}^2 \nonumber\\ &+ C_{bc} \Dot{\phi}_{C_{bc}}^2+C_{b} \Dot{\phi}_{b}^2+C_{bL} \Dot{\phi}_{C_{bL}}^2\Bigg]
\end{align}
and the potential is
\begin{align}\label{V}
    \mathcal{V} = \frac{1}{2}\Bigg[ \frac{\phi_{a}^2}{L_a}+\frac{\phi_{b}^2}{L_b}\Bigg] - 2E_{J}\cos(\frac{\pi\phi_{\mathrm{ext}}(t)}{\phi_0})\cos(\frac{2\pi\phi_{c}}{\phi_0}),
\end{align}
where the detail of the derivation of the coupler potential is provided in the appendix \ref{simplecoupler}.
Using Kirchoff's voltage laws 
\begin{align}\label{K}
    &\Dot{\phi}_{C_{ac}} = \Dot{\phi}_{c}-\Dot{\phi}_{a}
    \,,\,\,\,\, \Dot{\phi}_{C_{bc}} = \Dot{\phi}_{b}-\Dot{\phi}_{c}\,,\,\,\,\, \Dot{\phi}_{C_{bL}} = -\Dot{\phi}_{b},
\end{align}
and the conjugate relation
\begin{align}
    Q_i = \frac{\partial \mathcal{L}}{\partial \dot{\phi_{i}}}, i=a,c,b
\end{align}
the kinetic energy in the charge basis $[Q_a,Q_c,Q_b]$ is obtained as
\begin{align}\label{K2}
    \mathcal{T} = \frac{1}{2} \vec{Q}\mathcal{C}\vec{Q}^T 
\end{align}
where 
\begin{align}\label{J3}
    \mathcal{C}\approx
    \begin{bmatrix}
    \frac{1}{\mathbf{C_a}}&\,\,\,\,\frac{C_{ac}}{\mathbf{C_a}\mathbf{C_c}}&0\\\\
    \frac{C_{ac}}{\mathbf{C_a}\mathbf{C_c}}&\frac{1}{\mathbf{C_c}}&\frac{C_{bc}}{\mathbf{C_b}\mathbf{C_c}}\\\\
    0&\frac{C_{bc}}{\mathbf{C_b}\mathbf{C_c}}&\frac{1}{\mathbf{C_b}}
    \end{bmatrix}.
\end{align}
In the capacity matrix, $\mathbf{C_a} = C_a +C_{ac},\mathbf{C_c} =C_c+C_{bc}+C_{ac} , \mathbf{C_b} = C_b+C_{bc}+C_{bL}$. According to the capacitance matrix, the coupling strength between the coupler and the storage ( leakage ) cavity depends on $g_{ac} \propto \frac{C_{ac}}{\mathbf{C_a}\mathbf{C_c}},g_{bc} \propto \frac{C_{bc}}{\mathbf{C_b}\mathbf{C_c}}$, respectively, where we assume the coupling is weak $C_{ac},C_{bc}\ll C_{a},C_{b},C_{c}$. 

\subsection{Linear and nonlinear potential of the coupler}\label{L-nL Hamiltonian}
From Eq. \eqref{V}, the potential energy of the coupler includes the contribution of the nonlinear flux operators and the external drive as
\begin{align}
   U_{\mathrm{coupler}}= & -2E_{J}\cos(\frac{\pi\phi_{\mathrm{ext}}}{\phi_0})\cos(\frac{2\pi\phi_{c}}{\phi_0})\nonumber \\ =& -2E_{J}\cos(\frac{\varphi_{\mathrm{ext}}(t)}{2})\cos(\varphi_{c}),\label{a}
\end{align}
where the reduced flux is considered as $\varphi_c = \frac{2\pi \phi_c}{\phi_0}$ and $\varphi_{\mathrm{ext}}(t) = \varphi_{\mathrm{dc}} + F(t)\sin(\omega_d t).$ Using the Taylor expansion
\begin{align}
    \cos(\varphi_{c}) =& 1 - \frac{\varphi_{c}^2}{2} + \frac{\varphi_{c}^4}{4!}+ \mathcal{O} (\varphi_{c}^6) ,\label{potential2}
\end{align}
the potential energy of the transmon can be decomposed into a linear $U_C^l$ and nonlinear $U_C^{nl}$ terms as (the constant terms are dropped)
\begin{align}\label{nl1}
    U_{\mathrm{coupler}} =&\, U_C^l+U_C^{nl},
    \nonumber\\U_C^l =& -2E_J\cos(\frac{\varphi_{\mathrm{ext}}(t)}{2})[1-\frac{\varphi_{c}^2}{2}], \nonumber\\
    U_C^{nl}=&-2E_J\cos(\frac{\varphi_{\mathrm{ext}}(t)}{2})[\frac{\varphi_{c}^4}{4!} ].
    \end{align}
By expanding the flux drive term
\begin{align}\label{potential1}
    \cos(\frac{\varphi_{\mathrm{ext}}(t)}{2}) =& \cos(\frac{\varphi_{\mathrm{dc}}}{2}) - \frac{\sin(\frac{\varphi_{\mathrm{dc}}}{2}) F(t)\sin(\omega_d t) }{2} \nonumber\\&- \frac{\cos(\frac{\varphi_{\mathrm{dc}}}{2})}{8}F(t)^2\sin(\omega_d t )^{2} + \mathcal{O}(F(t)^3),
\end{align}
the quadratic part of the potential, second line of \eqref{nl1}, can be decomposed into time-independent and time-dependent terms as 
\begin{align}
    U_C^l =\,&E_J\Bigg[\cos(\frac{\varphi_{\mathrm{dc}}}{2})\Bigg]\varphi_c^2 \nonumber\\
   &-E_J\Bigg[\frac{\sin(\frac{\varphi_{\mathrm{dc}}}{2}) F(t)\sin(\omega_d t)}{2} \nonumber\\&\,\,\,\,\,\,\,\,\,\,\,\,\,\,\,\,\,+ \frac{\cos(\frac{\varphi_{\mathrm{dc}}}{2})}{8}F(t)^2\sin(\omega_d t)^{2}
 \Bigg]\varphi_c^2.\label{nl}
\end{align}
Consequently one can easily find the time-independent potential matrix $\mathcal{V'}$ in basis $\phi = (\phi_a,\phi_c,\phi_b)$ \begin{align}\label{J2}
     \mathcal{V'} = \begin{bmatrix}
    \frac{1}{L_A}&0&0\\\\
    0&\frac{8\pi^2 E_J(\cos(\frac{\varphi_{\mathrm{dc}}}{2}))}{\phi_0^2}&0\\\\
    0&0&\frac{1}{L_B}
    \end{bmatrix},
\end{align}
which makes the potential energy $\mathcal{V} = \frac{1}{2}\Vec{\phi}\mathcal{V'}\Vec{\phi}^T$.
Using the linear potential $\mathcal{V}$ and the Kinetic energy $\mathcal{T}$, the dressed mode of the circuit can be evaluated. In the dressed mode, the time-dependent and nonlinear terms of the potential of the coupler provide the value of the effective nonlinear interaction on the cavities and the optimal swap operator between the cavities, which are discussed in the next section.  %linear Hamiltonian and is employed to find the dressed modes of the circuit according to which the effective nonlinear interactions and the swap operator can be obtained which is discussed in the next section.

\subsection{Dressed modes and the effective Hamiltonian}\label{Cdressed}
To calculate the dressed mode of the circuit Fig. \ref{F1-cavitycoupler}, one can define the linear and time-independent Hamiltonian as
\begin{align}
    H^l = \frac{1}{2}\Vec{Q} \mathcal{C}\Vec{Q}^T+ \frac{1}{2}\Vec{\phi}\mathcal{V'}\Vec{\phi}^T
\end{align}
where $\mathcal{C},\mathcal{V'}$ are evaluated in Eqs. (\ref{J3},\ref{J2}), respectively.
The equation of motion for the Heisenberg operator is given by
\begin{align}
    \partial^2_t \Vec{Q}^T &= -\mathcal{V'C}\Vec{Q}^T\nonumber\\
    \partial^2_t \Vec{\phi}^T &= -\mathcal{CV'}\Vec{\phi}^T,
\end{align}
and can be solved by making the ansatz
\begin{align}
    \vec{Q}^T &= i\sum_n \sqrt{\frac{\hbar\omega_n}{2}}\frac{1}{\sqrt{\mathcal{C}}}\vec{\zeta}^T_n(a_n^\dagger e^{i\omega_n t}-a_n e^{-i\omega_n t })\nonumber\\
    \vec{\phi}^T &= \sum_n \sqrt{\frac{\hbar}{2\omega_n}}\sqrt{\mathcal{C}}\vec{\zeta}^T_n(a_n^\dagger e^{i\omega_n t}+a_n e^{-i\omega_n t }).
\end{align}
The frequency of the modes $\omega_n$ and its corresponding orthogonal mode-functions $\vec{\zeta}_n$ follow from the eigenvalue equation
\begin{align}
(\sqrt{\mathcal{C}}\mathcal{V'}\sqrt{\mathcal{C}} - \omega_n^2 \mathpzc{I})\vec{\zeta}^T_n = 0,
\end{align}
where yields the uncoupled Hamiltonian
\begin{align}\label{h}
    H^l/h = \sum_{n=1}^3 \omega_n a_n^\dagger a_n.
\end{align}
The phase operator of the coupler $\phi_c$, in the new basis $a_i$, can be expressed as
\begin{align}\label{dressed1}
   \phi_c  =& \sum_{n=1}^3 \frac{\lambda_{n}}{\sqrt{2}}(\hat{a}^\dagger_n + \hat{a}_n) = \sum_{n=1}^3\frac{A_n +A^{\dagger}_n}{\sqrt{2}} \nonumber \\ &\longrightarrow \lambda_{n} = \sqrt{\frac{\hbar}{\omega_n}}\vec{e}_c \sqrt{\mathcal{C}} \vec{\zeta}_n^T ,
\end{align}
where indices $i=1,2,3$ correspond to $a,c,b$, respectively and we introduce new parameters representation
\begin{align}
    \lambda_{a}&\equiv \lambda_1\,\,,\,\,\lambda_{c}\equiv\lambda_2\,\,,\,\,\lambda_{b}\equiv \lambda_3\nonumber\\ a &= a_1,c = a_2,b = a_3,\nonumber\\
    \omega_a &= \omega_1,\omega_c = \omega_2,\omega_b = \omega_3.
\end{align}
Considering a compact form of the flux of the tunable coupler \cite{nigg2012black,marcos2013superconducting}
\begin{align}\label{dressed}
\phi_{c} = \frac{1}{\sqrt{2}}(A_a+A_a^{\dagger}+A_b+A_b^{\dagger}+A_c+A_c^{\dagger}) = \frac{A+A^\dagger}{\sqrt{2}}
\end{align}
and defining $\bar{\lambda} = [A,A^\dagger] = |\lambda_{a}|^2+|\lambda_{b}|^2+|\lambda_{c}|^2$, 
the time-dependent parts of Eq. \eqref{nl} is obtained as 
\begin{align}\label{L1}
    H^l(t) = 
    \frac{-2E_J\pi^2}{\phi_0^2}\Bigg[&\frac{\sin(\frac{\varphi_{\mathrm{dc}}}{2}) F(t)\sin(\omega_d t)}{2} \nonumber\\ &+ \frac{\cos(\frac{\varphi_{\mathrm{dc}}}{2})}{8}F(t)^2\sin(\omega_d t)^{2}\Bigg](2A^\dagger A + \bar{\lambda}),
\end{align}
where provides the swap operator and the stark shifts. In addition, the fourth order of the flux operator, $U_C^{nl}$ in Eq. \eqref{nl1}, provides selfKerr, crossKerr interaction, and also Stark shifts which in the dressed mode of Eq. \eqref{dressed} is obtained
\begin{align}
   U_C^{nl} = H_{nl} \approx & \frac{-\pi^4E_J\cos(\frac{\varphi_{\mathrm{ext}}(t)}{2})}{3\phi_0^4}(A+A^\dagger)^4 \nonumber\\
   H_{nl} \approx &\frac{-\pi^4E_J\cos(\frac{\varphi_{\mathrm{ext}}(t)}{2})}{3\phi_0^4}\Bigg[12 \bar{\lambda} A^\dagger A +6 A^\dagger A^\dagger A A \Bigg].\label{NL}
\end{align}
If we consider the drive frequency $\omega_d = \omega_b - \omega_a$, the total Hamiltonian $ H =$ \ref{h}+\ref{L1}+\ref{NL}, in the rotating frame $\omega_{a}a^\dagger a +\omega_{b}b^\dagger b+\omega_{c}c^\dagger c$ is obtained
\begin{align}\label{Hamiltonian}
H =&  -i g_{\mathrm{swap}}(t)[b^\dagger a-a^\dagger b]\nonumber\\ &+ g_{Stark}(t)[\lambda_a^2 a^\dagger a+\lambda_b^2 b^\dagger b+\lambda_c^2 c^\dagger c]\nonumber\\
&+ g_{\mathrm{selfkerr}}[\lambda_a^4 a^{2\dagger}a^2+\lambda_b^4 b^{2\dagger}b^2+\lambda_c^4 c^{2\dagger}c^2]\nonumber\\
&+g_{\mathrm{crosskerr}}[\lambda_a^2 \lambda_b^2 b^\dagger ba^\dagger a+\lambda_a^2 \lambda_c^2 c^\dagger ca^\dagger a+\lambda_c^2 \lambda_b^2 b^\dagger bc^\dagger c],
\end{align}
where the coefficients are evaluated as
\begin{align}\label{Coeff}
    g_{\mathrm{swap}}(t) &= (\frac{\pi^2}{\phi_0^2}-\frac{\bar{\lambda}\pi^4}{\phi_0^4})E_J \lambda_a \lambda_b \sin(\frac{\varphi_{\mathrm{dc}}}{2})F(t),\\
    g_{\mathrm{Stark}}(t) &= (\frac{\bar{\lambda}\pi^4}{\phi_0^4}-\frac{\pi^2}{\phi_0^2})\frac{E_J \cos(\frac{\varphi_{\mathrm{dc}}}{2})}{4}F(t)^2,\\
 g_{\mathrm{selfkerr}}&=\frac{-2\pi^4E_J\cos(\frac{\varphi_{\mathrm{dc}}}{2})}{\phi_0^4}, \\
g_{\mathrm{crosskerr}}&=\frac{-8\pi^4E_J\cos(\frac{\varphi_{\mathrm{dc}}}{2})}{\phi_0^4}.\label{Coeff1}
\end{align} 

It is worth noting that because the coupler is initialized in the ground state, the Hamiltonian \eqref{Hamiltonian}, decouples the coupler from both cavities which means the terms $(c^\dagger c a^\dagger a, c^\dagger c b^\dagger b, c^{\dagger 2} c^2 )$ vanish at all times. Comparing Eqs. (\ref{H1},\ref{Hamiltonian}-\ref{Coeff1}), the following parameters are introduced
\begin{align}\label{Coeff2}
    \chi_{a(b)} =& \frac{-2\pi^4E_J\cos(\frac{\varphi_{\mathrm{dc}}}{2})}{\phi_0^4} \lambda_{a(b)}^4, \nonumber\\
    \chi_{ab} =& \frac{-8\pi^4E_J\cos(\frac{\varphi_{\mathrm{dc}}}{2})}{\phi_0^4} \lambda_a^2\lambda_b^2,\nonumber\\
    S_{a(b)}(t) =& g_{\mathrm{Stark}}(t) \lambda_{a(b)}^2.
\end{align}

%%%%%%%%%%%%%%%%%%%%%%%%%%%%%%%%%%%%
%%%%%%%%%%%%%%%%%%%%%%%%%%%%%%%%%%%%
%%%% Extra part of the appendix if we want to explain the calculation of the coupler ############################

\section{ Calculation of the potential of the coupler}\label{simplecoupler}
\begin{figure}[ht!]
\centering
\includegraphics[width=.32\textwidth]{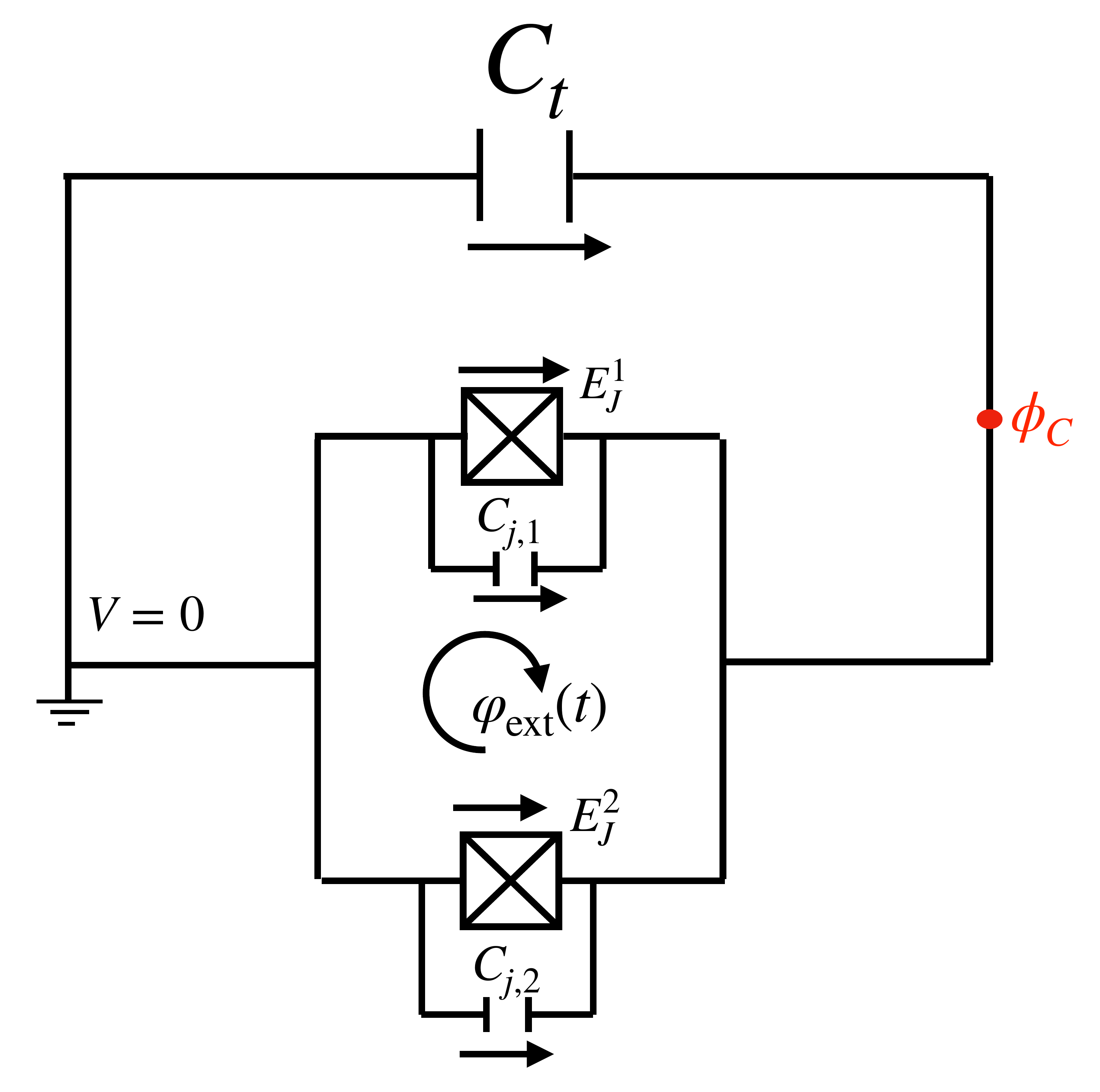}
\caption{Schematic of a flux-tunable coupler. Two Josephson junction $E_J^1,E_J^1$ are in parallel with two parasitic (small) capacitance $C_{j,1},C_{j,2}$ and shunted by a large capacitance $C_t$ \cite{rasmussen2021superconducting}. The flux drive in the small loop is $\varphi_{\mathrm{ext}}(t)$ and the bigger loop is described by the flux operator $\phi_C$.}
\label{SC}
\end{figure}

In general, we can consider each junction coupled in parallel with parasitic capacitance ($C_{j,1},C_{j,2}$), and the total system is shunted by a large capacitance $C_t$; see Fig. \ref{SC}. The Kinetic and potential energy of this simple circuit is
\begin{align}\label{V8}
    \mathcal{T} &= \frac{1}{2}\Bigg[ C_{t} \Dot{\phi}_{C}^2 + C_{j,1} \Dot{\phi}_{{j,1}}^2+C_{j,2} \Dot{\phi}_{{j,2}}^2  \Bigg]\nonumber\\
    \mathcal{V} &= - E_{j,1}\cos(\frac{2\pi\phi_{j,1}}{\phi_0})-E_{j,2}\cos(\frac{2\pi\phi_{j,2}}{\phi_0})
\end{align}
According to the number of nodes in the circuit, it can be described by one degree of freedom which we consider $\phi_{C}$.
Using the Kirchhoff voltage low, the small loop of junctions obeys the following equation
\begin{align}\label{V9}
    \phi_{j,1} - \phi_{j,2} = \phi_{\mathrm{ext}}.
\end{align}
In the following, we introduce the independent parameter $ \phi_{C} $ as the function of the two fluxes across the junctions
\begin{align}\label{V22}
    m_1 \phi_{j,1} + m_2 \phi_{j,2} = \phi_{C}
\end{align}
Using Eq. (\ref{V9},\ref{V22}) the fluxes are obtained as
\begin{align}
    \phi_{j,1} &= \frac{\phi_{C}+m_2 \phi_{\mathrm{ext}}}{m_1+m_2}\nonumber\\
    \phi_{j,2} &= \frac{\phi_{C}-m_1 \phi_{\mathrm{ext}}}{m_1+m_2},
\end{align}
where subsequently, the kinetic part of Eq. \eqref{V8} can be rewritten as
\begin{align}
   \mathcal{T} = \frac{1}{2}\Bigg[& [\frac{C_{j,1}+C_{j,2}}{(m_1+m_2)^2}+C_t]\Dot{\phi}_{C}^2\nonumber\\&+\frac{2(m_2 C_{j,1}-m_1 C_{j,2})}{(m_1+m_2)^2}\Dot{\phi}_{C}\Dot{\phi}_{\mathrm{ext}} + \mathcal{O}(\Dot{\phi}_{\mathrm{ext}})^2\Bigg].
\end{align}
We do not take the second order of the flux fluctuation into account, as we assume the flux drive is slow in time.
To remove the effect of the flux drive fluctuation $\Dot{\phi}_{\mathrm{ext}}$, one can utilize the following two conditions 
\begin{align}\label{V44}
    m_1+m_2 &= 1\nonumber\\
    m_2 C_{j,1}-m_1 C_{j,2} &= 0,
\end{align}
where leads to
\begin{align}\label{V55}
    m_1 &= \frac{C_{j,1}}{C_{j,1}+C_{j,2}}\nonumber\\
    m_2 &= \frac{C_{j,2}}{C_{j,1}+C_{j,2}}.\nonumber
\end{align}
Hence $\mathcal{T}, \mathcal{V}$ are obtained as
\begin{align}
      \mathcal{T} &= \frac{C_{j,1}+C_{j,2}+C_{t} }{2}\Dot{\phi}_{C}^2 \nonumber\\
       \mathcal{V} &= -E_{J,1}\cos(\frac{2\pi[\phi_{C}+m_2 \phi_{\mathrm{ext}}]}{\phi_0})-E_{J,2}\cos(\frac{2\pi[\phi_{C}-m_1 \phi_{\mathrm{ext}}]}{\phi_0}).
\end{align}
If we consider symmetry junction with $E_{J,1} = E_{J,2} = E_J$ and $C_{j,1} = C_{j,2}$, consequently $m_1=m_2=1/2$ and the potential energy can be simplified as
\begin{align}
    \mathcal{V} = -2 E_{j} \cos(\frac{2\pi\phi_{C}}{\phi_0})\cos(\frac{\pi \phi_{\mathrm{ext}}}{\phi_0}),
\end{align}
which is equivalent to Eq. \eqref{V}. Introducing $C_c \equiv C_{j,1}+C_{j,2}+C_{t} $, the same kinetic energy as in Equation \eqref{T} is obtained.

\section{Wigner Function of Fig. \ref{Fig4} (c)}\label{SC1}
\begin{figure*}[ht!]
\centering
\includegraphics[width=1\textwidth]{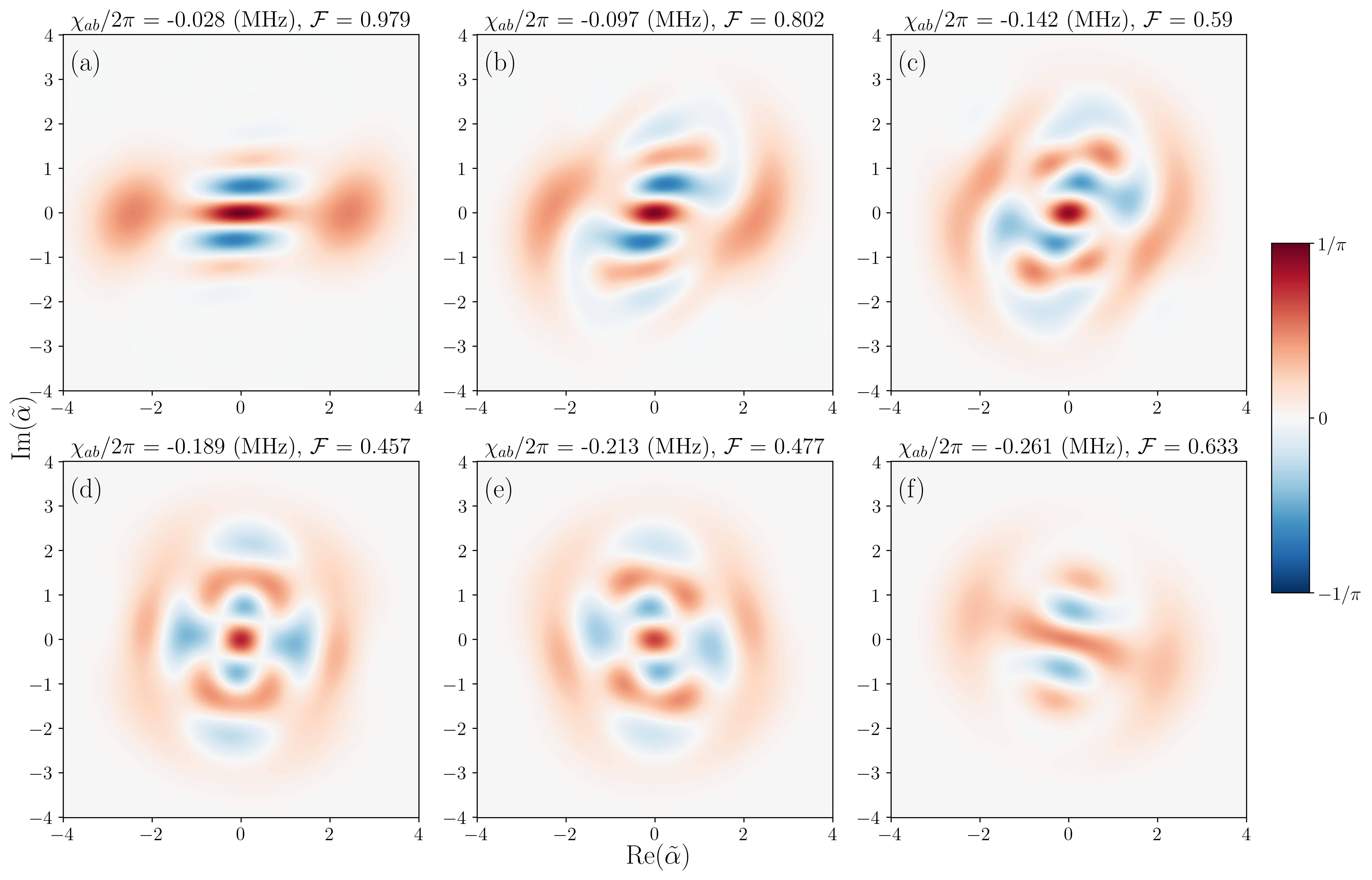}
\caption{Different realizations of the Wigner functions of the single output mode TCCS in Fig. \ref{Fig4} (c). }
\label{SC}
\end{figure*} 
\bibliographystyle{unsrt}
%\bibliography{Ref}

\end{document}